\documentclass[a4paper,fleqn]{article}
\usepackage[dvips]{graphics}
\usepackage{graphicx}
\usepackage{amsmath}   
\usepackage{amssymb}
\usepackage[usenames]{color}
\usepackage{multirow}

\usepackage[margin=1.2in]{geometry}

\numberwithin{equation}{section}

\newcommand{\la}{\left\langle}
\newcommand{\ra}{\right\rangle}

\newcommand{\lla}{\langle\!\langle}
\newcommand{\rra}{\rangle\!\rangle}
\newcommand{\p}{\partial}

\begin{document}

\title{Retarded action principle and self-financing portfolio dynamics}
\date{\today}
\author{D Lesnik}
\maketitle


\begin{abstract}
We derive a consistent differential representation for the dynamics of a self-financing portfolio for different hedging strategies. In the basis of the derivation there is the so called ``retarded action principle'', which represents the causality in the evolution of dependent stochastic variables. We demonstrate this principle on example of a vanilla and a storage option. 
\end{abstract}

\section{Introduction}

In the first part of this paper we derive an evolution equation for a self-financing portfolio consisting of a vanilla option, a dynamically traded linear hedge and a cash account. The portfolio is self-financing in a sense that it can be considered as ``isolated'' -- the cash flow is solely generated by trading the hedge, and there is no external source of cash required.

We use the evolution equation to investigate the dynamics of the portfolio with different hedging strategies. We will be interested in the terminal portfolio distribution. Obviously, the option pricing formula does not effect the terminal portfolio value, since at the maturity it simply becomes equal to the deterministic pay-off. On the contrary, different hedging strategies effect the distribution and the expected value of the terminal portfolio:
\begin{itemize}
	\item If a trader holds a naked (unhedged) vanilla option, the terminal value of the portfolio is random, and the expected portfolio value equals the expected pay-off, i.e. the probabilistic option price. 
    \item The so called ``risk neutral'' hedging strategy makes the terminal portfolio equal the ``risk neutral'' option price with arbitrarily narrow distribution. The ``risk neutral'' portfolio is insensitive to the market drift, which makes it so special.
    \item An arbitrary hedge strategy on a drift-less market leads to the expected terminal portfolio value which coincides with the ``risk neutral'' portfolio, but the distribution depends on the hedging strategy.
    \item On the market with drift the expected value and the distribution of the terminal portfolio depend on the hedging strategy.
\end{itemize}

For our analysis we consider three different option pricing formulas -- Black-Scholes (we will also refer to the BS option price and hedging strategy as ``risk-neutral''), probabilistic and intrinsic. Apart from that we look at different hedging strategies, and in particular, we consider delta hedging corresponding to different option pricing formulas. We also show how the option time value can be calculated from the intrinsic hedging strategy.

In the second part we generalise the equations for the storage option, and show how the storage option time value can be calculated by following the rolling intrinsic hedging strategy.

\section{Market with drift}

Let $F(t)$ be the price of the underlying at time $t$. The underlying is supposed to follow a generalised Wiener process with drift:
\begin{align}
    \label{eq:price_process}
    & dF(t,T) = \mu(F, t)\,dt + \sigma(F,t)\,dW_t\,.
\end{align}
For geometric Brownian motion (GBM) we have $\mu = \mu_0\,F$, $\sigma = \sigma_0\,F$. The drift term is supposed to be real and statistically significant, unlike the virtual drift considered in risk neutral pricing theory (see Sec. \ref{sec:virtual_drift}).

The presence of the drift affects the out-turn distribution, and introduces a statistical arbitrage opportunities. We will use the following identities:
\begin{align}
    &\la dF \ra = \mu(F,t)\,dt\\
    & dF^2 = \sigma^2(F,t)\,dt
\end{align}
An important aspect of the market with drift is that a consistent and statistically significant drift can only be observed on markets incomplete with limited liquidity. On the other hand we will be dealing with the portfolio evolution equation, which is derived under the assumption, that the trader performs a continuous dynamic hedging, which is hardly possible on the illiquid markets. Hence, the results for the markets with drift should be considered as an approximation, which can only be approached in reality to the extent allowed by the market liquidity.

A drift-less market implies $\mu= 0$, and hence $\la dF\ra = 0$. For the formulas which are derived under assumption of the drift-less market we will use the label
\[
    [\text{no drift}]:
\]

\subsection{Virtual drift and risk neutral pricing}
\label{sec:virtual_drift}

A standard option pricing theory is developed under the assumption of effective market hypothesis, as well as market completeness, i.e. liquidity and absence of market friction. Under such conditions any stochastic model of the asset price dynamic must obey one crucial property: the corresponding price process should be a martingale Markov process. Any deviation from martingale property leads immediately to the statistical arbitrage, which is impossible within the framework of efficient liquid market.

It has been observed for a large class of assets that a typical behaviour of the asset price has three major dynamic characteristics -- drift, diffusion and defaults. The drift characterises a short term systematic drift of the price, diffusion describes a random zero-mean fluctuations, defaults describe a random occurrence of price drops. For some assets the event of default means jump to zero, i.e. the asses ceases existing afterwards. For other assets the default could be only partial, allowing some ``recovery rate''. The simplest price process capturing all three features could be represented in the following form
\begin{align}
    \frac{dF}{F} = \tilde \mu\,dt + \sigma\,dW + dZ
\end{align}
where $\tilde \mu$ describes the short-term drift, $dW$ describes the diffusion, and $dZ$ is a stochastic jumps process, describing the random default events. The martingale condition yields:
\begin{align}
    \tilde \mu\,dt + \la dZ\ra = 0
\end{align}
Thus the drift term is not arbitrary, but is determined by the defaults jump process. This relation reflects the well know fact that riskier assets (i.e. those assets having bigger default rate) have higher short-term growth rate.

It is common in some financial literature to consider the \emph{truncated} price process, in which the jumps component is dropped:
\begin{align}
    \label{eq:truncated_process}
    \frac{dF}{F} \cong \tilde \mu\,dt + \sigma\,dW 
\end{align}
In such truncated form the equation is violating the martingale condition (to emphasise this we used the symbol $\cong$). Despite this shortcoming, in some circumstances the latter model can be useful. Often the defaults process has very long expectation times between jumps. In such situation Eq.~(\ref{eq:truncated_process}) gives a good description of the price process dynamics on short times. The advantage of this equation lies in its practicality, since it is easily tractable analytically, whereas most of results based on the jump-diffusion process are either very complicated, or do not have closed form solutions at all.

However, as has been noticed, Eq.~(\ref{eq:truncated_process}) violates the martingale condition, and as a results, cannot be used for probabilistic option pricing. Indeed, the expected price increment becomes
\[
    \la dF\ra \cong \tilde\mu\,F\,dt
\]
which leads to a systematic bias in the option price. The drift term $\tilde \mu$ only makes sense if it is compensated by the defaults jump process. We refer to the value $\tilde \mu$ as \emph{virtual drift}, to emphasise that it represents an uncompensated drift, which violates the martingale condition. 

The distribution of the out-turn price following the truncated evolution equation is called ``real world measure''. Ironically it is not describing the measure of the real world, since it contains a systematic bias.

In order to deal with the inconsistent price dynamics, a risk-neutral pricing theory has been developed. This theory uses the portfolio replication approach (delta-hedging strategy) in order to derive (or better to say -- define) the risk neutral option price, which turns out to be independent on the virtual drift. The risk neutral price is equivalent to the probabilistic price under the simplified price process 
\begin{align}
    \frac{dF}{F} = \sigma\,dW 
\end{align}
This model does not have any short-term drift or defaults jump component. However it preserves the martingale condition, and in many cases is sufficient for option pricing. Rather sophisticated risk-neutral pricing theory, a great deal of which is based on measure theory, is essentially needed to justify why the virtual drift term should be omitted.

In our paper we sometimes consider a price processes with drift, such as in Eq.~(\ref{eq:price_process}). We should emphasise that the drift $\mu$ in this case is not virtual drift, but a true observable and statistically significant drift. As has been noticed previously, the drift is only possible in markets with limited liquidity, which does not allow the majority of market participants to take and advantage of the statistical arbitrage. Market with limited liquidity does not have to satisfy the martingale condition. The option pricing in this situation becomes rather ambiguous, since it depends on the hedging strategy. Whatever the hedging strategy is chosen, we use the probabilistic approach to derive the option price.

\section{Averaging}

We will be dealing with averaging over the stochastic price process. In this regard we need to make some comments about the notation.

Let $f(t, F)$ be some function of the price process $F(t)$ and time. An average of $f$ is an expected value of $f$ as a function of random variable $F(T)$, given the initial condition $F(0)$. We denote this average as
\[
    \la f \ra_F
\]
which is short for $\la f(t, F(t))|\mathcal{F}_0\ra_F$. Here $\mathcal{F}_t$ is a filtration generated by Brownian motion.

A differential increment $dF$ can be expressed in terms of $F$ and the Brownian driver differential $dW$. We can define an averaging over $dW$ under filtration $\mathcal{F}_t$. We designate the average over $dW$ as
\[
    \la dF \ra_{dW}
\]
which is short for $\la dF(F, dW) | \mathcal{F}_t \ra_{dW}$. A differential of any explicit function of $F$ becomes a function of $dW$, so we designate the average over $dW$ as
\[
    \la df \ra_{dW}
\]

Notice that under filtration $\mathcal{F}_0$ the differential $df(t, F)$ is a function of two random variables -- $F(t)$ and $dW$, and hence the average $\la df(t)\ra_{dW}$ is a random variable. Thus for differential expressions we can introduce a full average, which includes averaging over the increment $dW$ and over the stochastic variable $F(t)$.
\[
    \la df \ra_{F, dW} = \la \la df\ra_{dW}\ra_F
\]
As an example consider a differential form $h\,dg$, where $h(t,F)$ is a function of the stochastic process $F$, and $dg$ is a function of $F$ and $dW$. When averaging over $dW$, the function $h$ can be taken out of the averaging bracket:
\[
\la h\,dg\ra_{dW} = h\,\la dg\ra_{dW}
\]
For the full average we get:
\[
    \la h\,dg\ra = \la h\,\la dg\ra_{dW}\ra_{F}
\]
In this case the averaging over $F$ cannot be factorised, as both values $h$ and $\la dg\ra_{dW}$ are functions of $F$. In particular
\[
    \la h\,dF\ra = \la h\, \mu \ra_{F}\;dt
\]

Notice yet another useful expression valid for the drift-less price process:
\[
[\text{no drift}]:\quad\la h\,dF\ra = \la h\,\la dF\ra_{dW}\ra_{F} = 0
\]

We will omit the subscript for the averaging bracket, if it is obvious from the context what is meant. Usually we will mean a full average for the expressions like $\la g\ra$ or $\la h\,dg\ra$, and an average over $dW$ for the expressions like $\la dg\ra$ or $\la dF\ra$.

\section{Portfolio value function}
\label{sec:portfolio_value}

Let us consider a portfolio consisting of a vanilla option expiring at some time $T_e$, a linear hedge position $h(t)$ and a cash account~$P(t)$. We designate $H(t)$ the value of the hedge position. 

Let also $C(F, t)$ be the option price. It should be stressed, that $C(F, t)$ does not imply any particular pricing formula, e.g. Black-Scholes. The only requirement to the option price is that at maturity it coincides with the option pay-off 
\[
    C(F, T_e) = K(F)
\]
As discussed in the introduction, we consider three different special cases of the option price -- risk-neutral, intrinsic and probabilistic. 

The portfolio value is defined as
\begin{align}
    & \Pi(t) = C(t) + H(t) + P(t)
\end{align}
It is important to understand that our definition of the portfolio value is not the same as the portfolio market price, i.e. in general case $\Pi(t)$ is not the price at which the portfolio could (or should) be bought or sold on the market. Indeed, even in a liquid drift-less market we know that a correct market price of portfolio is obtained by using BS option pricing formula. The intrinsic call option price disregards the option time value, and hence the corresponding portfolio value underestimates its market price. However at the time of option expiry all option pricing formulas agree, since they coincide with the option pay-off. The terminal portfolio value is thus independent on the option pricing formula. It simply equals the sum of the option pay-off and the cumulative cash flow from the hedge trades.

In illiquid markets with drift the notion of portfolio market price is not well defined at all. Firstly, the price is not observed on the market because the market is illiquid. The price is not well defined mathematically either. Indeed, the terminal portfolio distribution, -- mean and variance, -- depend on the hedging strategy. A risk neutral strategy has vanishing risk (there is always some residual risk caused by market incompleteness and lack of liquidity). However there exist other strategies, that lead to a higher expected profit, which comes in a package with an irreducible risk. A trader with some risk appetite might consider the risky strategy more attractive. Thus the ``fair'' price doesn't exist, as different traders who have different risk profile, have different view on the fair price. In contrast, the portfolio value $\Pi(t)$ is well defined. The terminal portfolio distribution provides a detailed information about the expected profit and associated risk, and so we will be interested in analysing the terminal portfolio value distribution. In particular, the expected terminal portfolio value will be our main focus.

Let $f(F,t)$ be the risk-neutral, $g(F,t)$ be the probabilistic, and $I(F)$ be the intrinsic option price. We consider three definitions of the portfolio: ``risk-neutral'' 
\begin{align}
    \Pi_{rn} = f(F, t) + H(t) + P(t)
\end{align}
``probabilistic'' 
\begin{align}
    \Pi_{p} = g(F, t) + H(t) + P(t)
\end{align}
and ``intrinsic''
\begin{align}
    \Pi_{i} = I(F) + H(t) + P(t)
\end{align}
In all three cases considered above the initial portfolio value is different, and defined by the initial option price
\begin{align}
    & H(0) = 0\\
    & P(0) = 0 \\
    & \Pi(0) = C(F_0, 0)
\end{align}

The terminal option value does not depend on the pricing formula, as it coincides at expiry with the pay-off $K(F)$ 
\[
    C(F,T_e) = f(F,T_e) = g(F,T_e) = I(F,T_e)= K(F)
\]
The expected terminal option value is nothing else but the expected pay-off, and hence coincides with the initial probabilistic option value
\begin{align}
    \la C(F, T_e)\ra = \la K(F)\ra = g(0)
\end{align}
which is valid for any option price formula, e.g. 
\[
    \la f(T_e)\ra = \la g(T_e)\ra = \la I(T_e)\ra = g(0)
\]
Terminal portfolio value depends on the terminal option price and on the cash flow generated by the hedge trades. Terminal portfolio value is a random variable (except for the risk-neutral hedge strategy). So we can speak of the terminal portfolio distribution. We designate
\[
    \Pi_e = \Pi(T_e)
\]
This value is correctly defined, as $\Pi(T_e)$ doesn't depend on the option pricing formula. In particular, we will be interested in the expected value $\la \Pi_e\ra$.

As will be shown below, in the drift-less market the expected value $\la \Pi_e\ra$ does not depend on the hedging strategy, and coincides with the risk-neutral option value. The standard deviation on the other hand depends on the hedging strategy. The only risk-neutral strategy leads to zero standard deviation.

In the market with drift both expected value $\la \Pi_e\ra$ and standard deviation $\lla \Pi_e^2\rra$ are affected by the hedging strategy. Risk-neutral hedging strategy gives rise to risk-neutral terminal value and zero standard deviation. Other hedging strategies give rise to different expected value and standard deviation of $\Pi_e$.

\section{Dynamics of the portfolio components}

In this section we consider the evolution equation governing the portfolio dynamics. 

Everywhere below we assume zero interest rate, which is equivalent to using a risk-less bond as a numeraire. A transition to a currency units can be made by a change of variables (see App.~\ref{app:ir}).

\subsection{Hedge value dynamics}

The value of the hedge is defined as
\begin{align}
    H(t) = h(t)\,F(t)
\end{align}
where $F$ is the spot price, and $h$ is the hedge position. For simplicity of notation we omit the time argument where it does not lead to an ambiguity. Both $h$ and $F$ are stochastic variables. The differential increment of the hedge is given by the Ito rule
\begin{align}
    dH = h\,dF + F\,dh + dh\,dF = h\,dF + (F+dF)\,dh\,.
\end{align}
The first expression in the right hand side is symmetric with respect to both variables, but the second expression can be given a meaningful interpretation. The term $h\,dF$ reflects the change of the hedge value due to the change in the forward price. The second term $(F + dF)\,dh$ corresponds to the additional value associated with updating the hedge position. Since the new hedge is calculated after the price change is observed, the correction of the hedge position $dh$ has to be purchased at the new price $F+dF$.

\subsection{Cash account dynamics and retarded action principle}

The dynamics of the cash account results from the changing hedge position. The differential increment of the cash account is given by
\begin{align}
    \label{eq:retarded_action_principle}
    dP = -(F+dF)\,dh
\end{align}
We refer to this relation as ``\emph{retarded action}'' principle. It reflects the fact that the hedge position is updated after the change of price is observed. Retarded action principle is a differential representation of the \emph{self-financing} portfolio.\footnote{Retarded action principle reflects causality. The market price increment $dF$ is by definition an independent stochastic variable. On the contrary, the hedge increment $dh$ is defined by the hedge strategy, and hence is a function of time and market observables: $dh_t = dh(t, F, dF)$. Thus, between the price increment $dF$ and hedge increment $dh$ there is a causal relationship -- the latter is a function of the former, but not the other way round. However, an observer who is given a sequence of observations of $dF$ and $dh$ for successive time moments, would not be able to tell which of the variables is a function of another. Symmetric expression of the hedge value differential $dH$ contains no information about the cause and effect relationship between $dF$ and $dh$. The expression for $dP$ is in turn asymmetric, and reflects the causality: the increment $dh$ is calculated after the increment $dF$ is observed, and this leads to appearance of an additional term $dF\,dh$ in the cash flow differential. Notice that the causality is only a feature of the Wiener process, since the additional term in the cash flow is quadratic, and would be disregarded in an ordinary differential calculus.}

The expression (\ref{eq:retarded_action_principle}) is derived in such a way that the hedge portfolio does not change due to the change of hedge position. Indeed, the combination of the hedge with the cash account has the following dynamics
\begin{align}
    \label{eq:dH+dP}
    d(H+P) = h\,dF
\end{align}
This expression does not contain the term proportional to $dh$, which can be considered as a formal definition of retarded action principle. Eq.~(\ref{eq:dH+dP}) in particular implies that opening or closing the hedge position has no impact on the portfolio, even if $dh$ is not infinitesimally small.

Notice that in a drift-less market the hedge on average earns no money:
\begin{align}
    \text{[no drift]:}\quad \la d(H+P) \ra_{dW} = h\,\la dF \ra_{dW} =  0 
\end{align}
The relation between expected hedge and cash flow increments
\begin{align}
    \text{[no drift]:}\quad \la dH \ra_{dW} = - \la dP \ra_{dW}
\end{align}
will be used later in application to storage option.

In presence of the market drift the expected drift of the hedge portfolio does not vanish, and gives a non-trivial contribution to the entire portfolio evolution.

\subsection{Option price dynamics}
Using Ito rule we can write
\begin{align}
    dC = \dot C\,dt + C'\,dF + \frac12\,C''\,dF^2
\end{align}
where we designated
\[
    \dot C = \frac{\p C}{\p t}\,,\quad
    C' = \frac{\p C}{\p F}\,,\quad
    C''= \frac{\p^2 C}{\p F^2}
\]
This evolution equation is valid for any option pricing formula, provided it can be expressed as a function of $t$ and $F$. Below we will consider special cases of risk neutral, probabilistic and intrinsic option prices.

\subsection{Full portfolio dynamics}

Taking into account Eq.~(\ref{eq:dH+dP}), the full portfolio dynamics becomes
\begin{align}
    \label{eq:dPi_1}
    d\Pi = dC + h\,dF
\end{align}
Substituting the expression for the option differential, we obtain
\begin{align}
    \label{eq:dPi_2}
    d\Pi = \dot C\,dt + (C' + h)\,dF + \frac12\,C''\,dF^2
\end{align}
As discussed before, the initial portfolio value coincides with the option price 
\[
    \Pi(0) = C(0) 
\]
This value is correctly defined in a sense that setting up of an initial hedge has zero cost:
\[
    H(+0) + P(+0) = h(0)\,dF = 0
\]
and hence the portfolio experiences no discontinuity at the first time moment.

The terminal portfolio value is obtained by integrating Eq.~(\ref{eq:dPi_1}) over the time:
\begin{align}
    \Pi_e = C(T_e) + \int_0^{T_e} h\,dF
\end{align}
In the drift-less market the average of the second term vanishes, and we obtain for the expected value
\begin{align}
    \text{[no drift]:}\quad & \la d\Pi(t)\ra_{dW} = \la dC \ra_{dW} \\
    \label{eq:expected_term_value}
                            & \la \Pi_e\ra = \la C(T_e) \ra = g(0)
\end{align}
The expected terminal portfolio value is solely determined by the option pay-off and by the terminal distribution of the underlying.

In the market with drift the term $\la h\, dF\ra$ does not vanish, so the expected terminal portfolio value becomes
\begin{align}
    \la \Pi_e \ra = \la C(T_a)\ra + \int \la h\,dF\ra = 
    g(0) + \int \la h\, dF\ra
\end{align}
Thus the terminal portfolio value depends on the hedging strategy. Below we consider a few examples.

\subsection{Delta-hedged portfolio dynamics}

A delta hedging strategy is defined as
\begin{align}
    h(t) = -C'(t)
\end{align}
Notice that this strategy depends on how the option price is defined. Below we consider three different delta-hedging strategies -- \emph{risk-neutral} (corresponding to the risk-neutral option price), \emph{drift-adjusted} (corresponding to the probabilistic option price) and \emph{intrinsic} (corresponding to the intrinsic option price).

We will use ``hat'' to denote the special case of delta hedged portfolio:
\[
    \hat \Pi
\]
Delta hedged portfolio obeys the following dynamics
\begin{align}
    \label{eq:dPi_delta_hedged}
    d\hat \Pi = \dot C\,dt + \frac12\,C''\,dF^2
\end{align}
The increment of the delta hedged portfolio does not depend on $dF$, and hence is deterministic under filtration $\mathcal{F}_t$.\footnote{We point out that the object $dF^2$ can formally be considered as deterministic, which is a jargon adopted in Ito calculus. In the discrete time for any arbitrarily small but fixed time increment $dt$ the value $dF$ is a random number, and so is $dF^2$. However in the continuous time framework the standard deviation of the integral $\int dF^2$ vanishes in the limit of $dt\to 0$. For this reason we can formally think of $dF^2$ as being deterministic.}  Notice that the delta hedged portfolio is deterministic regardless of the option pricing formula. This is a consequence of our definition of the portfolio value (see Sec.~\ref{sec:portfolio_value}).

\section{Hedging strategies}

\subsection{Risk-neutral delta hedging}

For the ``risk neutral'' option price the portfolio dynamics is given by
\[
d\Pi_{rn} = df + h\,dF = \dot f\,dt + (f' + h)\,dF + \frac12\,f''\,dF^2
\]
The risk neutral option value $f(F,t)$ can be derived according the following logic. Let us use a risk neutral delta hedging strategy
\[
    h = -f'
\]
Then the portfolio dynamics becomes
\begin{align}
    d\hat\Pi_{rn} = \dot f\,dt + \frac12\,f''\,dF^2
\end{align}
This expression is deterministic (i.e. risk-less) under filtration $\mathcal{F}_t$. The ``no free lunch'' argument of the risk neutral pricing theory is that the risk-less portfolio should be constant. Thus the risk neutral option price should obey the relation
\begin{align}
    \label{eq:BS76}
    \dot f\,dt + \frac12\,f''\,dF^2=0
\end{align}
known as Black Scholes equation.

If the option price is defined by the BS equation, and if one follows the delta hedging strategy, then the portfolio value is constant:
\[
    d\hat\Pi_{rn}\equiv 0\,,\qquad \hat\Pi_{rn}(T_e) = \hat\Pi_{rn}(0)\,.
\]
This is the remarkable property of the risk neutral pricing theory: by delta hedging the portfolio can be made risk-less and independent on the market drift. Another property of the risk neutral hedging is that the hedge replicates the option evolution. Indeed, from $d\Pi_{rn} = df + dH + dP = 0$ we get
\begin{align}
    \label{eq:replication}
    dH + dP = -df
\end{align}
which means that the hedge part of the portfolio replicates the behaviour of the option with the opposite sign. 

It is worth emphasising that for a finite time step $\Delta t$ the left hand side of Eq.~(\ref{eq:BS76}) is a zero-mean stochastic variable
\[
    \dot f\,\Delta t + \frac12\,f''\,\Delta F^2 = \varepsilon\,,\qquad 
    \la \varepsilon\ra = 0
\]
and hence the replication formula~(\ref{eq:replication}) is only valid up to the random variable $\varepsilon$. More accurate replication can be achieved by hedging with non-linear products that would replicate option gamma $f''$ and option theta $\dot f$.

Noticing that $\Pi_{rn}(0) = f(0)$ and on the drift-less market $\la\Pi_e\ra = g(0)$, we conclude
\[
    \text{[no drift]:} \quad g(0) = f(0)
\]
Thus the risk neutral option price coincides with the probabilistic price if the market dynamics is drift-less.

Let us now consider a portfolio which uses BS option price but an arbitrary hedging strategy. Taking into account BS equation, the option price evolution can be represented as
\begin{align}
    \label{eq:df}
    df = f'\,dF
\end{align}
and hence for an arbitrary hedging strategy we obtain: 
\begin{align}
    \label{eq:dPrn}
    d\Pi_{rn} = (f' + h)\,dF
\end{align}
Integrating over the time and averaging yields
\[
    \la \Pi_e\ra = \Pi_{rn}(0) + \int_0^{T_e}\la (f' + h)\, dF\ra =
    f(0) + \int_0^{T_e} \la (f' + h)\,\mu \ra\;dt
\]
Averaging the squared expression for $\Pi_e$ we obtain
\begin{align*}
    & \lla \Pi_e^2 \rra = 
    \int_0^{T_e} \lla x(t_1)\,x(t_2)\,\mu(t_1)\,\mu(t_2)\rra\;dt_1\,dt_2 + 
    \int_0^{T_e} \lla x^2\,\sigma^2\rra\;dt \\
    &\text{where}\qquad  x = f' - h
\end{align*}
The expressions for $\la \Pi_e\ra$ and $\lla \Pi_e^2 \rra$ allow to calculate the influence of an arbitrary hedging strategy on the terminal portfolio distribution in presence of the market drift. These expressions, simple though they may look, involve averaging over the price path scenarios $F(t)$, which requires $h$ to be expressed as an explicit function of $F$. If $h$ is a functional on $F$, then the averaging over the realisations of the price paths will involve path integrals, and the entire calculation may become highly non-trivial.

As an example of a slightly more interesting hedging strategy let's consider a bid-offer optimised hedging. This hedging strategy is a rule at every time step to only partially update the hedge position in order to balance between the risk of unhedged part of the portfolio and the cost of market friction. This strategy could be modelled by the following equation
\[
    dh = -k\,(h + f')\,dt
\]
with some constant $k$. Obviously
\[
    h(t) = \begin{cases}
    h(0)\,\quad & \text{for}\quad  k = 0 \\
    -f'\, & \text{for}\quad  k \to \infty
    \end{cases}
\]
Integrating over time we obtain
\[
    h(t) = h(0)\,e^{-k\,t} - k \int_0^t e^{-k(t-u)}\,f'(u)\,du
\]
The option delta $f'$ can be explicitly expressed as a function of $F$
\[
    f'(t) \equiv f'(t, F(t))
\]
Thus the hedge position becomes a functional $h[F(t)]$. Estimation of the terminal portfolio distribution for this hedging strategy would require calculation of path integrals.

\subsection{Drift-adjusted delta hedging}

We start with the differential expression for the probabilistic option value $g$:
\begin{align}
    \label{eq:dg1}
    dg = \dot g\,dt + g'\,dF + \frac12 \,g''\,dF^2
\end{align}
The probabilistic option value has a property of zero expected drift
\begin{align}
    \la dg \ra = 0
\end{align}
This property follows from the probabilistic option price definition. Indeed, for the two arbitrary time moments $t_1$ and $t_2>t_1$ we have
\begin{align*}
    g(t_1) = \la g_e|F_1\ra & = \int dF_e\; P(F_e|F_1) \, K(F_e) = 
    \int dF_2 \; P(F_2|F_1)\int dF_e\; P(F_e|F_2)\,K(F_e) = \\
    & = \la g(t_2)|\mathcal{F}_{t_1}\ra
\end{align*}
where $\la x|y\ra$ is an expectation of $x$ conditional on $y$, $P(x_1|x_0)$ is a probability distribution function of $x_1$ conditional on $x_0$. We also designated $g_e = g(T_e), F_e = F(T_e), F_1 = F(t_1), F_2 = F(t_2)$. From the latter equation follows $\la dg\ra$ = $\la g(t+dt)\ra - g(t) = 0$.

By averaging the expression for $dg$ over $dW$, we find
\begin{align}
    \label{eq:BS_for_g}
    \dot g\,dt + g'\,\la dF\ra + \frac12 \,g''\,dF^2 = 0
\end{align}
In the drift-less market $\la dF\ra=0$, and $g$ obeys the same BS equation with the same terminal boundary condition as the risk neutral option price. Thus, as expected, the probabilistic option price coincides with the risk neutral price.

From Eqs.~(\ref{eq:dg1}, \ref{eq:BS_for_g}) we obtain
\begin{align}
    dg = g'\,(dF - \la dF\ra)
\end{align}
(compare this with Eq.~(\ref{eq:df})). The portfolio dynamics follows as
\begin{align}
    d\Pi_p = (g' + h)\,dF - g'\,\la dF\ra
\end{align}
Averaging this expression over $dW$ yields
\[
    \la d\Pi_p\ra = h\la dF\ra
\]
Hence the probabilistic portfolio has a drift, and there exists no hedging strategy (apart from a trivial one) that would eliminate the drift.\footnote{The portfolio has an irreducible drift because we used the probabilistic option price as part of the portfolio. This portfolio has a different starting point, and an irreducible drift if compared to the risk neutral portfolio.} 

Drift-adjusted delta hedging strategy is defined as
\begin{align}
    h = -g'
\end{align}
It gives rise to the following hedged portfolio dynamics
\begin{align}
    d\hat\Pi_p = - g'\,\la dF\ra
\end{align}
Similarly to the risk neutral portfolio, the delta hedged portfolio increment is deterministic. But unlike the risk neutral one, the probabilistic portfolio has a drift. The ``no free lunch'' argument does not apply here for the following reason. For the risk-neutral portfolio dynamics the option price followed from the ``no free lunch'' argument. However in case of drift-adjusted portfolio the option pricing formula is been postulated (which is in line with our portfolio value definition).

The terminal value of the delta hedged portfolio is given by
\begin{align}
    \hat\Pi_e = \hat\Pi_p(0) - \int_0^{T_e}  g'\,\la dF\ra = 
    \hat\Pi_p(0) - \int_0^{T_e}  g'\,\mu\;dt
\end{align}
This is a stochastic value, since both $g'$ and $\la dF\ra$ can be expressed as explicit functions of the price $F$. The expectation value of the terminal portfolio is obtained by averaging over the price path scenarios.

In case of an arbitrary hedging strategy the portfolio increment averaged over $dW$ is given by
\[
    \la d\Pi_p \ra_{dW} = h\,\la dF \ra = h\,\mu\;dt
\]
Thus the expectation value of the terminal portfolio is given by
\begin{align}
    \la \Pi_e \ra = \Pi_p(0) + \int_0^{T_e} \la h\,dF\ra = 
    \Pi_p(0) + \int_0^{T_e} \la h\,\mu\ra\,dt
\end{align}
where the averaging is performed over the price scenarios. Depending on the drift direction and hedging strategy, the expected terminal portfolio value can be made arbitrary large. A pitfall of this feature is that the terminal portfolio is a random number. Portfolio possesses an irreducible risk, which can only be minimised by approaching to the risk neutral hedging strategy, and giving up the opportunity of a higher profit.

\subsection{Rolling intrinsic delta hedging}

Intrinsic option price is characterised by no explicit time dependence. The evolution of the option price is thus given by
\[
    dI = I'\,dF + \frac12\,I''\,dF^2
\]
Substituting the latter equation into the formula~(\ref{eq:dPi_2}) we obtain
\begin{align}
    d\Pi_i = (I'+h)\,dF + \frac12\,I''\,dF^2
\end{align}
Intrinsic delta hedging strategy is defined as
\[
    h = -I'
\]
The evolution of the delta hedged portfolio becomes
\[
    d\hat\Pi_i = \frac12\,I''\,dF^2 
\]
i.e. is driven by the intrinsic option gamma. Integrating $d\hat\Pi_i$ over the time and averaging yields
\begin{align}
    \la \hat\Pi_e\ra - \hat\Pi_i(0) =  \frac12 \int_0^{T_e} \la I''\,dF^2 \ra
\end{align}
The initial value of the portfolio equals initial intrinsic option value $I(0)$. The expected value of the terminal portfolio $\la \hat\Pi_e\ra$ under the risk neutral measure equals an expected option pay-off (see Eq.~(\ref{eq:expected_term_value})), which is equal to the risk neutral option price. Thus the risk neutral option time value can be calculated as 
\begin{align}
    \text{[no drift]:}\quad V_T = 
    \frac12 \int_0^{T_e} \la I''\,dF^2 \ra = 
    \int_0^{T_e} \la dI \ra
\end{align}
In the last integral the averaging is full, i.e. over $dW$ and $F$.

\subsubsection{Call option time value}

Here we suppose that the market has no drift $\la dF\ra=0$.

Let us consider a european call option. The intrinsic option price is given by
\begin{align}
    I(F) = (F-K)\,\theta[F-K]
\end{align}
where $\theta[x]$ is the Heaviside function, $F$ is the spot price and $K$ is the strike. The variation of the intrinsic option price is easy to calculate:
\begin{align}
    dI = \theta[F-K]\,dF + \frac12\,\delta[F-K]\,dF^2
\end{align}
where $\delta[x]$ is the Dirac delta function. The first and second order terms give option delta and gamma:
\begin{align}
    I' = \theta[F-K]\,,\qquad I'' = \delta[F-K]
\end{align}
The dynamics of the intrinsically hedged portfolio becomes
\begin{align}
    d\hat\Pi_i = \frac12\,\delta[F-K]\,dF^2\,.
\end{align}

Notice that for the rolling intrinsic strategy the portfolio value is strictly growing. This result is known by traders, and is sometimes used as a conservative trading strategy, since PnL (though $\hat\Pi_i$ underestimates the market value of the portfolio) is never decreasing. 

By integrating $d\hat\Pi_i$ over time and averaging, we obtain the option time value as
\begin{align}
    \label{eq:intrinsic_time_value}
    V_T = \frac12 \int_0^{T_e} \la \delta[F-K]\,dF^2 \ra 
\end{align}
We perform the explicit calculation for demonstration purpose in appendix~\ref{app:call_option_time_value}.

\section{Generalisation for a storage option on a drift-less market}

Let $t$ be the observation time, and $T$ be the delivery time. Let $F_t(T)$ be the forward price observed on time $t$ with delivery on time $T$. $F_t(T)$ as a function of $T$ is called forward curve.

In this section we consider a drift-less market. The forward curve is supposed to follow a martingale continuous stochastic process with independent increments, for instance
\[
    \frac{dF_t(T)}{F_t(T)} = \sigma\, dW_t\,,\qquad \la dF_t(T)\ra \equiv 0
\]
Particular form of the process does not play a role. Here $dF_t(T)$ is a variation associated with the time increment $dt$ and considered as a function of $T$. Generally for any function $f_t(T)$ we will be using variation denoted as $df_t(T)$ or $\delta f_t(T)$ and defined as an increment over the time interval $dt$:
\[
    \delta f_t(T) = f_{t+dt}(T) - f_t(T)
\]

Let $T_e$ be the storage contract duration, such that 
\[
    t\in [0, T_e]
\]
On every time moment $t$ the hedge position is a function of delivery time:
\[
    h_t(T)
\]
The value of the hedge is defined as
\begin{align}
    H(t) = \int_t^{T_e} h_t(T)\,F_t(T)\,dT
\end{align}
This definition implies that the value of the linear forward contract at time $t$ with delivery at time $T$ is assigned to the current time $t$. In reality the actual cash flow for the future contract may take place at some other time. However it is irrelevant from mathematical perspective. The latter definition is chosen for convenience.

We agree that by definition the initial hedge equals zero:
\[
    h_0(T)\equiv 0\,,\qquad H(0) = 0
\]

The increment of the hedge value is given by
\begin{align}
    dH(t) = -h_t(t)\,F_t(t)\,dt + 
    \int_t^{T_e} \Big(h_t(T)\,\delta F_t(T) + 
    \Big(F_t(T) + \delta F_t(T)\Big)\,\delta h_t(T)\Big)\,dT
\end{align}
where $\delta F_t(T)$ and $\delta h_t(T)$ are the variations corresponding to the observation time increment $dt$. The first term is nothing else but the exercise trade -- the prompt part of the hedge portfolio which becomes physical. The value of the hedge becomes smaller by the exercise trade due to the shortening time horizon. The exercise trade should also be reflected in the changing option value.

The increment of the cash account according to the retarded action principle is given by
\begin{align}
    \label{eq:dP_storage}
    dP(t) = - \int_t^{T_e} \Big(F_t(T) + 
    \delta F_t(T)\Big)\,\delta h_t(T)\,dT
\end{align}
Notice that there is no cash flow associated with the exercise trade, as the corresponding underlying has been paid for on the future market during the hedging. 

We define the exercise trade as
\begin{align}
    dE(t) = h_t(t)\,F_t(t)\,dt\,,\qquad E(0) = 0
\end{align}
where $F_t(t)$ is the spot price on time $t$. 

Let $C(t)$ be the price of the storage option for the remaining time horizon $[t, T_e]$ calculated with initial conditions on $t$. For the entire portfolio increment we thus obtain:
\begin{align}
    \label{eq:dPi_storage}
    d\Pi(t) = dC(t) - dE(t) + \int_t^{T_e} h_t(T)\,\delta F_t(T)\,dT
\end{align}
Comparing the latter expression with Eq.~(\ref{eq:dPi_1}) we notice that the storage portfolio dynamics has one additional term $dE$.

The initial portfolio value coincides with the initial option price
\begin{align}
    \Pi(0) = C(0)
\end{align}
This is obvious since setting up of the initial hedge costs nothing:
\[
    H(+0) + P(+0) = -h_0(0)\,F_0(0)\,dt + 
    \int_0^{T_e} [h_0(T)\,\delta F_0(T)]\,dT = 0
\]
where we used $h_0(T)\equiv 0$. The portfolio experiences no jump due to the setting up of the initial hedge, and hence is correctly defined at time $t=0$.

The terminal option value is obviously zero
\[
    C(T_e) = 0
\]
Now averaging Eq.~(\ref{eq:dPi_storage}) and integrating over the time yields
\begin{align}
    \label{eq:Expected_Pi(Te)}
    \la \Pi_e\ra = -\int_0^{T_e} \la dE\ra = -\la E(T_e)\ra
\end{align}
where we made use of the drift-less market assumption. The conclusion from this expression is the following. The expected terminal portfolio value is solely defined by the exercise trades, and is not affected by the hedge trades. Hedge only has an impact on the distribution of the terminal portfolio around the expected value. For a storage option there exists an optimal exercise strategy, which maximises the expected option value. A suboptimal exercise strategy leads to an irreducible loss of the expected option value.

Next we introduce the target function
\begin{align}
    & S(t) = C(t) - E(t) 
\end{align}
with boundary conditions
\begin{align}
    & S(0) = C(0) \\
    & \la S(T_e)\ra = -\la E(T_e)\ra = \la \Pi_e\ra
\end{align}
With this definition the portfolio dynamics can be written as
\begin{align}
    \label{eq:dPi_storage2}
    d\Pi(t) = dS(t) + \int_t^{T_e} h_t(T)\,\delta F_t(T)\,dT
\end{align}
which is the storage option generalisation of Eq.~(\ref{eq:dPi_1}). Using drift-less assumption, we find
\begin{align}
    \la d\Pi\ra_{dW} = \la dS \ra_{dW}
\end{align}

The target function is an ordinary function of time $t$, and a functional on the forward curve $F_t(T)$, and hence the full differential can be expressed by a generalised Ito formula
\begin{align}
    \label{eq:dS_generic}
    dS(t) = \frac{\p S}{\p t}\,dt +
    \int_t^{T_e}\frac{\delta S}{\delta F(u)}\,\delta F(u)\,du 
    + \frac12\, \iint_t^{T_e} \frac{\delta^2 S}{\delta F(u)\,
        \delta F(v)}\,\delta F(u)\,\delta F(v)\,du\,dv
\end{align}
In the latter formula and below we omit the subscript $t$ for simplicity of notation, e.g. $\delta F(u)$ implies $\delta F_t(u)$. Combining the expression for $dS$ with Eq.~(\ref{eq:dPi_storage2}) and introducing the correlation function
\begin{align}
    \Lambda_t(u,v) = \frac{1}{dt}\,\delta F_t(u)\,\delta F_t(v)
\end{align}
we obtain for the portfolio dynamics
\[
    d\Pi = \frac{\p S}{\p t}\,dt +
    \int_t^{T_e}\left(\frac{\delta S}{\delta F(u)} + h(u)\right)\,\delta F(u)\,du 
    + \frac{dt}{2}\, \iint_t^{T_e} \frac{\delta^2 S}{\delta F(u)\,
        \delta F(v)}\,\Lambda(u,v)\,du\,dv
\]
This formula is a storage option generalisation of the portfolio evolution equation~(\ref{eq:dPi_2}) for a vanilla option. Here $\p S/\p t$ represents option Theta, functional derivative $\delta S/\delta F(u)$ represents option Delta, and second functional derivative $\delta^2 S/ \delta F(u)\,\delta F(v)$ represents option Gamma.

Delta hedge is defined as
\[
    h_t(T) = -\frac{\delta S}{\delta F_t(T)}
\]
This is a continuous time analogue of the vanilla option delta hedge. For the dynamics of the delta-hedged portfolio we obtain
\begin{align}
    d\hat\Pi = \frac{\p S}{\p t}\,dt +
    \frac{dt}{2}\, \iint_t^{T_e} \frac{\delta^2 S}{\delta F(u)\,
        \delta F(v)}\,\Lambda(u,v)\,du\,dv
\end{align}
This is an analogue of Eq.~(\ref{eq:dPi_delta_hedged}).

\subsection{Risk neutral pricing}

Suppose that $C(t)$ represents a risk neutral storage option price over the time horizon $[t, T_e]$. In this case we can follow the same logic as for the risk neutral vanilla option pricing. Since the delta hedged portfolio increment is deterministic (under filtration $\mathcal{F}_t$), it is not exposed to the market risk, and hence should be constant. We thus obtain
\begin{align}
    \frac{\p S}{\p t} +
    \frac12\, \iint_t^{T_e} \frac{\delta^2 S}{\delta F(u)\,
        \delta F(v)}\,\Lambda(u,v)\;du\,dv = 0
\end{align}
This is a contingent claim generalisation of Black Scholes equation. Formally this is an integro-differential equation on an infinitely-dimensional function $S$, considered as a function of $t$ and $F_t(T)$ for all possible values of $T$. At any time $t$, given the forward curve $F_t(T)$, functional $S[t, F_t(T)]$ equals the storage option value, in a similar way as the function $f(t,F(t))$ equals a vanilla option price for the given time $t$ and spot price $F(t)$. Noticing that the time derivative $\p S/\p t$ is the storage option theta, whereas the second variational derivative $\delta^2 S/\delta F(u)\,\delta F(v)$ represents the storage option gamma, the BS equation establishes a link between theta and gamma in exactly the same way it does for a vanilla option.

Taking BS equation into account, the delta hedged portfolio dynamics becomes trivial
\[
    d\hat\Pi \equiv 0
\]
For an arbitrary hedging strategy the portfolio dynamics becomes
\begin{align}
    d\Pi = \int_t^{T_e}\left(\frac{\delta S}{\delta F(u)} + h_t(u)\right)\,\delta F(u)\,du 
\end{align}
This equation allows to estimate the portfolio drift and variance
\begin{align*}
    & \la d\Pi\ra = 0 \\
    & d\Pi^2 = dt \iint_t^{T_e} 
        \left(\frac{\delta S}{\delta F(u)} + h_t(u)\right)\,
        \left(\frac{\delta S}{\delta F(v)} + h_t(v)\right)\,
        \Lambda(u,v)\; du\,dv
\end{align*}
provided $S$ can be expressed explicitly as a functional on $F_t(T)$.

\subsection{Rolling intrinsic strategy}

In this section we consider a rolling intrinsic storage option pricing and hedging strategy.

The rolling intrinsic strategy is defined as follows. For every observation time $t$ there is a forward curve $F_t(T)$ observed on the market. If the storage level at time $t$ is $q(t)$, then we can solve an intrinsic optimisation problem for the remaining time horizon $[t, T_e]$ with initial condition $q(t)$. Let us designate the intrinsic optimal exercise strategy calculated at time $t$ as $\dot q_{int}(t,T)$. We interpret here $q(t,T)$ is a function of time $T$, parametrised by the observation time $t$. The time derivative is taken with respect to $T$: $\dot q_{int}(t, T) = \p q_{int}(t,T)/\p T$. For simplicity of notation we will designate
\[
    r_t(T) = \dot q_{int}(t,T)
\]
The intrinsic storage option value can be expressed simply as 
\begin{align}
    \label{eq:full_intrinsic_target_function}
    I(t) = - \int_t^{T_e} r_t(T)\,F_t(T)\,dT
\end{align}
This integral represents the future profit calculated with the current forward curve on intrinsic exercise profile. 

According to the rolling intrinsic strategy the hedge position at time $t$ is defined as
\[
    h_t(T) = -\frac{\delta I}{\delta F_t(T)} = r_t(T)
\]
and the exercise trade
\[
    dE(t) = r_t(t) \, F_t(t)\, dt
\]

For the correct option pricing it is crucial that the exercise trade is optimal. It has been shown in~\cite{2015arXiv150606979L} that the optimal storage exercise coincides in the leading order with the intrinsic exercise. This justifies using the rolling intrinsic strategy for estimating the expected storage option value.

The initial value of $I$ provides the storage option intrinsic value
\[
    I(0) = V_{int}
\]
The terminal value of $I$ is obviously zero:
\[
    I(T_e) = 0
\]
The exercise $E(t)$ is given by integrating the exercise trades
\[
    E(t) = \int_0^t r_\tau(\tau)\, F_\tau(\tau)\,d\tau
\]
This integral represents the value of the closed trades. Here we used that the initial value of $E$ is zero $E(0) = 0$. The expected terminal value of $E$ represents the expected option value with the opposite sign (see Eq.~(\ref{eq:Expected_Pi(Te)})).
\[
    -\la E(T_e) \ra = \la \Pi_e\ra
\]
Now for the target function $S = I - E$ we have
\begin{align}
    & S(0) = V_{int} \\
    & \la S(T_e)\ra = \la \Pi_e\ra
\end{align}
Hence, the time value can be obtained as
\begin{align}
    V_T = \int_0^{T_e} \la dS \ra_{F, dW}\,,
\end{align}
Next we notice that the target function $S$ depends on $t$ only through the limits of the integration in $E$ and $I$. It is easy to see that
\[
    \frac{\p I}{\p t}  = \frac{\p E}{\p t} = r_t(t)\,F_t(t)
\]
and hence
\[
    \frac{\p S}{\p t} = 0
\]
Finally we obtain the evolution equation for the target function
\begin{align}
    dS(t) = \int_t^{T_e}\frac{\delta S}{\delta F(u)}\,\delta F(u)\,du 
    + \frac{dt}{2}\, \iint_t^{T_e} \frac{\delta^2 S}{\delta F(u)\,
        \delta F(v)}\,\Lambda(u,v)\;du\,dv
\end{align}
Now averaging and taking into account that $\la \delta F\ra \equiv 0$ we obtain
\[
    \la dS(t)\ra = 
    \frac{dt}{2}\, \iint_t^{T_e} \la \frac{\delta^2 S}{\delta F(u)\,
        \delta F(v)}\,\Lambda(u,v)\ra\,du\,dv
\]
The time value follows as
\begin{align}
    V_t = \int_0^{T_e} dt\, 
    \iint_t^{T_e} \frac12 \la \frac{\delta^2 S}{\delta F(u)\,
        \delta F(v)}\,\Lambda(u,v)\ra\,du\,dv
\end{align}
This expression allows to calculate the time value if the target function can be expressed as an explicit functional of the forward curve $F(T)$. 

We can obtain another expression for the time value which contains the variation of the optimal intrinsic strategy $\delta r_t(T) = r_{t+dt}(T) - r_t(T)$. In order to do so, we notice that the ``closed'' part $E(t)$ of the target function is constant with respect to the forward curve variation (the variation of the forward curve impacts only the future values of the curve $F(t,T)$ for $T>t$), and hence
\begin{align}
    dS =  \eth I\,,
\end{align}
where we designated $\eth$ the variation which should be calculated for constant time $t$. Thus
\begin{align}
    V_T = \int_0^{T_e} \la \eth I \ra\,.
\end{align}
Next we observe that the hedge value coincides with the negative intrinsic option value:
\[
    H(t) = \int_t^{T_e} r_t(T)\,F_t(T)\,dT = -I(t)\,,
\]
and hence
\begin{align}
    \la \eth I \ra = -\la \eth H \ra = \la dP \ra
\end{align}
where $P$ is the cumulative cash flow from the hedge trades. The second relation follows from the definitions of $H$ and $dP$, if we take into account that $\la \delta F\ra \equiv 0$ and that the variation $\eth$ keeps the integration limits constant.

Thus the time value can be calculated as
\[
    V_T = \int_0^{T_e} \la dP \ra\,.
\]
It can also be shown by direct calculation, that the expected cumulative cash flow from the hedge trades yields the option time value (see Appendix~\ref{app:Cumulative_cash_flow}).

Substituting the definition of $dP$:
\begin{align*}
    dP(t) = - \int_t^{T_e} ( F + \delta F)\,\delta r\;dT
\end{align*}
we finally obtain the time value as
\begin{align}
    V_T = - \int_0^{T_e} \, \left[ \int_t^{T_e} \la ( F + \delta F)\,\delta r \ra\,dT \right]
\end{align}
where the averaging is full. This expression allows to calculate the storage option time value, provided we know an analytic expression for the optimal intrinsic exercise strategy $r_t(T)$. Generally, the exercise strategy $r_t(T)$ is functional on the forward curve $F_t(T)$. The variation of the intrinsic exercise $\delta r$ should then be expressed in terms of the forward curve variation $\delta F$. After averaging, the first order terms with respect to $\delta F$ vanish. Thus, the time value will be expressed in terms of the correlation function $\Lambda_t(u,v)$.


\pagebreak

\appendix
\begin{large}APPENDICES\end{large}

\section{Change of variables for interest rate}
\label{app:ir}

Everywhere in the main text the prices are expressed in units of a risk-less bonds, and hence no interest rate is present in any evolution expression. In order to change to the units of a currency whose value decreases at the rate $r$, we need to introduce the following change of variables (we use ``bar'' to denote new variables):
\begin{enumerate}
    \item All values having units of currency, e.g. spot price $F$, option price $f$, etc., are scaled as
    \begin{align}
        & \bar F = e^{r\,t}\,F \\
        & \bar f = e^{r\,t}\,f
    \end{align}
    \item Considering the time as one of the independent variables, we formally introduce a new time $\bar t = t$. The partial derivatives are replaced according to
    \begin{align}
        & \frac{\p}{\p t} = \frac{\p}{\p\bar t} + r\,\bar F\,\frac{\p}{\p\bar F} \\
        & \frac{\p}{\p F} = e^{r\,\bar t}\,\frac{\p}{\p\bar F} \\
        & \frac{\p^2}{\p F^2} = e^{2\,r\,\bar t}\,\frac{\p^2}{\p\bar F^2}
    \end{align}
\end{enumerate}

\section{Vanilla Call Option Time Value}
\label{app:call_option_time_value}

We depart from the formula~(\ref{eq:intrinsic_time_value})
\begin{align}
    V_T = \frac12 \int_0^{T_e} \la \delta[F-K]\,dF^2 \ra
\end{align}
Let $F(t)$ follow GBM:
\begin{align}
    \frac{dF(t)}{F(t)} = \sigma\,dW_t\,;
\end{align}
The square of this expression yields
\begin{align}
    dF^2 = \sigma^2\,F^2\;dt\,.
\end{align}
and thus
\[
\Gamma = \frac{1}{dt} \la dI \ra_{F,dW} = 
\frac{\sigma^2}{2} \la \delta[F-K]\; F^2 \ra_F
\]
Now to calculate value $\Gamma$, we need to average this expression over the stochastic price $F$. Let $P(F,t)$ be the probability density function of $F$ at time $t$. For GBM it reads
\begin{align}
    P(F,t) = \frac{1}{F}\,\frac{1}{\sigma\,\sqrt{2\,\pi\,t}}\,\exp\left(
        -\frac{ \left(  \ln\frac{F}{F_0} + \frac{\sigma^2}{2}\,t \right)^2  }{2\,\sigma^2\,t}
        \right)\,;
\end{align}
For $\Gamma$ we obtain
\begin{align}
    \Gamma = \int P(x,t)\, \frac{\sigma^2\,x^2}{2}\,\delta[x-K] \; dx = 
    \frac{\sigma^2\,K^2}{2}\,P(K,t)\,;
\end{align}
Substituting $P$ we finally obtain
\begin{align}
    \Gamma = \frac{\sigma\,K}{2}\,\frac{1}{\sqrt{2\,\pi\,t}}\,\exp\left(
        -\frac{ \left(  \ln\frac{K}{F_0} + \frac{\sigma^2}{2}\,t \right)^2  }{2\,\sigma^2\,t}
        \right)\,;
\end{align}

The time value $V_T$ is obtained by integrating $\Gamma$ over the time horizon:
\begin{align}
    V_T = \int_0^{T_e} \Gamma\,dt
\end{align}

\subsection{Integration of $\Gamma$}

The change of variables
\begin{align}
    y = \sqrt{t}\,;\quad dy = \frac{dt}{ 2\,\sqrt{t}}
\end{align}
gives
\begin{align}
    \label{eq:VT_integral}
    V_T = \frac{\sigma\,K}{\sqrt{2\,\pi}}\,e^{-2\,a\,b}\,
    \int_0^{\sqrt{T}} \exp[-a^2\,y^2 - b^2/y^2]\,dy
\end{align}
where
\begin{align*}
    a = \frac{\sigma}{2\,\sqrt{2}}\,;\qquad 
    b = \frac{1}{\sigma\,\sqrt{2}}\,\ln\frac{K}{F_0}
\end{align*}

Making use of the integral from the Sec.~\ref{sec:integral} we obtain:
\begin{align}
    & F_0 < K \,:\qquad V_T = \frac{\sigma\,K}{\sqrt{2\,\pi}}\,e^{-2\,a\,b}\, J_+[\sqrt{T}] =
    \frac12\,\left(
        F_0 - K + F_0\,\Phi[k_1] + K\,\Phi[k_2]
        \right) \\
    & F_0 > K \,:\qquad V_T = \frac{\sigma\,K}{\sqrt{2\,\pi}}\,e^{-2\,a\,b}\, J_-[\sqrt{T}] =
    \frac12\,\left(
        K- F_0 + F_0\,\Phi[k_1] + K\,\Phi[k_2]
        \right)
\end{align}
where $\Phi[x]$ is the Error function, and
\begin{align*}
        k_1 = \frac{\frac{\sigma^2}{2}\,T - \ln\frac{K}{F_0}}{\sigma\,\sqrt{2\,T}}\,, \qquad
        k_2 = \frac{\frac{\sigma^2}{2}\,T + \ln\frac{K}{F_0}}{\sigma\,\sqrt{2\,T}}
\end{align*}

It is easy to show that the time value obtained by this method coincides with the Black-Scholes time value:
\begin{align*}
    V_T = CallOptionValue(F_0,K,T) - \max(F_0-K,0)\,.
\end{align*}

Note that the Black-Scholes option value is usually expressed in terms of the cumulative normal distribution function $N[x]$, which is related to the Error function as
\begin{align*}
    N[x] = \frac12\,\Big(1+ \Phi\left[\frac{x}{\sqrt{2}}\right] \Big)\,.
\end{align*}

\subsection{Integral evaluation}
\label{sec:integral}

We have an indefinite integral $I$ and definite version of it $J$:
\begin{align}
    I = \int e^{-a^2\,x^2 - b^2 / x^2}\,dx\,,\qquad
    J[T] = \int_0^T e^{-a^2\,x^2 - b^2 / x^2}\,dx
\end{align}

We are searching the solution for~(\ref{eq:VT_integral}) in the form of Error function:
\begin{align}
    \Phi(x) = \frac{2}{\sqrt{\pi}}\int_0^x e^{-t^2}\,dt\,,\qquad
    \Phi(\pm\infty) = \pm 1\,.
\end{align}

Noticing that
\begin{align}
        &\frac{d}{dx}\, \Phi\left(a\,x + b/x \right) = 
        \frac{2}{\sqrt{\pi}}\,e^{-2\,a\,b}\,
        \left( a-\frac{b}{x^2} \right)\,
        e^{-a^2\,x^2 - b^2 / x^2} \\
        &\frac{d}{dx}\, \Phi\left(a\,x - b/x \right) = 
        \frac{2}{\sqrt{\pi}}\,e^{2\,a\,b}\,
        \left( a+\frac{b}{x^2} \right)\,
        e^{-a^2\,x^2 - b^2 / x^2}
\end{align}
we find
\begin{align}
    &\frac{d}{dx}\,\left[ e^{4\,a\,b}\, \Phi\left(a\,x + b/x \right) + \Phi\left(a\,x - b/x \right)
    \right]
    = \frac{4\,a}{\sqrt{\pi}}\,e^{2\,a\,b}\, e^{-a^2\,x^2 - b^2/x^2}
\end{align}

Thus we obtain
\begin{align}
        I = C + \frac{\sqrt{\pi}}{4\,a}\,\Big(
            e^{2\,a\,b}\,\Phi(a\,x+b/x) + e^{-2\,a\,b}\,\Phi(a\,x-b/x)
        \Big)\,;
\end{align}

Using the limit for $T\to 0$
\begin{align}
    & b>0\,:\qquad \lim_{T\to +0} \Phi(a\,T + b/T) = 1\,,\qquad
    \lim_{T\to +0} \Phi(a\,T - b/T) = -1 \\
    & b<0\,:\qquad \lim_{T\to +0} \Phi(a\,T + b/T) = -1\,,\qquad
    \lim_{T\to +0} \Phi(a\,T - b/T) = 1
\end{align}
we find the definite integral:
\begin{align}
    & b>0\,:\quad J_+[T] = \frac{\sqrt{\pi}}{4\,a}\,\Big(
    e^{-2\,a\,b} - e^{2\,a\,b} + e^{2\,a\,b}\,\Phi(a\,T+b/T) + e^{-2\,a\,b}\, \Phi(a\,T - b/T)
    \Big) \\
    & b<0\,:\quad J_-[T] = \frac{\sqrt{\pi}}{4\,a}\,\Big(
    e^{2\,a\,b} - e^{-2\,a\,b} + e^{2\,a\,b}\,\Phi(a\,T+b/T) + e^{-2\,a\,b}\, \Phi(a\,T - b/T)
    \Big)
\end{align}

\section{Cumulative cash flow}
\label{app:Cumulative_cash_flow}

Here we show how the time value of a storage option can be obtained from the cumulative cash flow of rolling intrinsic hedge trades. Using Eq.~(\ref{eq:dP_storage}) we find the total cumulative cash flow as
\begin{align}
        P(T_e) = -\int_0^{T_e} dT \int_0^T (F_t(T) + 
        \delta F_t(T))\,\delta h_t(T)
\end{align}
where $h_t(T) = \dot q_{int}(t,T)$. The inner integral is over the observation time $t$. It gives a cumulative cash flow for the fixed delivery time $T$. 

Now we use integration by parts, which in our case reads
\begin{align}
    \left. (F\,h)\right|_{t=0}^T = 
    \int_0^T h\,\delta F + \int_0^T F\,\delta h + 
    \int_0^T \delta F\,\delta h
\end{align}
where the integration and limits are with respect to the time $t$. Averaging the latter expression and noticing that $\la \delta F\ra = 0$ we obtain
\[
    \la F_T\,h_T \ra - F_0\,h_0 = \int_0^T \la F\,\delta h \ra +
    \int_0^T \la \delta F\,\delta h \ra 
\]
The averaged cumulative cash flow becomes
\begin{align}
    \la P(T_e)\ra = \int_0^{T_e} F_0(T)\,h_0(T)\,dT -
    \int_0^{T_e} \la F_T(T)\,h_T(T)\ra \,dT
\end{align}
Noticing that 
\[
   - \int_0^{T_e} \dot q_{int}(0,T)\,F(0,T)\,dT
\]
equals the option intrinsic value, and 
\[
    -\la \int_0^{T_e} \dot q_{int}(T,T)\,F(T,T)\,dT \ra
\]
equals the true option value\footnote{We make use of the fact that the intrinsic exercise $\dot q_{int}(T,T)$ is optimal stochastic exercise.}, we conclude that
\begin{align}
    V_T = \la P(T_e) \ra = \int_0^{T_e} \la dP \ra\,.
\end{align}

\end{document}